\documentclass[aps,prl,reprint,groupedaddress]{revtex4-1}
\usepackage{graphicx}
\usepackage{amsmath}
\usepackage[T1]{fontenc}
\usepackage{epstopdf}

\begin{document}


\title{Perturbative solution to the SIS epidemic on networks}


\author{Lloyd P. Sanders}
\email[Corresponding author: ]{lloyd.sanders@thep.lu.se}
\altaffiliation{Department of Astronomy and Theoretical Physics, Lund University, SE-223 62 Lund, Sweden}

\author{Bo S\"oderberg}
\altaffiliation{Department of Astronomy and Theoretical Physics, Lund University, SE-223 62 Lund, Sweden}

\author{Dirk Brockmann}
\altaffiliation{Northwestern Institute on Complex Systems and Department of Engineering Sciences and Applied Mathematics, Northwestern
University, Evanston, IL, USA}

\author{Tobias Ambj\"ornsson}
\altaffiliation{Department of Astronomy and Theoretical Physics, Lund University, SE-223 62 Lund, Sweden}



\begin{abstract}
Herein we provide a closed form perturbative solution to a general $M$-node network SIS model using the transport rates between nodes as a perturbation parameter. We separate the dynamics into a short-time regime and a medium/long-time regime. We solve the short-time dynamics of the system and provide a limit before which our explicit, analytical result of the first-order perturbation for the medium/long-time regime is to be employed. These stitched calculations provide an approximation to the full temporal dynamics for rather general initial conditions.

To further corroborate our results, we solve the mean-field equations numerically for an infectious SIS outbreak in New Zealand (NZ, \emph{Aotearoa}) recomposed into 23 subpopulations where the virus is spread to different subpopulations via (documented) air traffic data, and the country is internationally quarantined. We demonstrate that our analytical predictions compare well to the numerical solution.
\end{abstract}


\maketitle


\section{Introduction}
In mathematical epidemiology the canonical deterministic susceptible-infected-susceptible (SIS) model is one of the most elementary compartmental models, receiving consistent attention (the specifics of which are given at length below) since the seminal work of Kermack and McKendrick \cite{Kermack1927}. Put simply, this model partitions a large, well mixed, homogeneous population into two compartments: susceptible and infected, where birth and death are neglected. A susceptible may become infected upon contact with another infected with some finite probability, and conversely an infected will recover after some typical time, becoming once more susceptible.

The simplicity of this model and its ability to characterize the main motifs of viral infections, where recovery does not assure immunity (for example Gonorrhea \cite{Hethcote1989} or Chlamydia \cite{Turner2006}), has allowed for extensive research. Due to the mathematical tractability of the mean-field model, many extensions have been applied, to include other important dynamical factors \cite{Llibre2008, Das2010, Mendez2012}. Although these models have focused on deterministic mean-field approaches as we shall herein, there is also a surge in converting the models over to their stochastic counterparts \cite{Gillespie2007}, and analyzing different properties of the system (for a recent example see \cite{Keeling2008}).

Recently, network theory \cite{Newman2003} has sought to understand the large scale realism of human mobility \cite{Brockmann2011, Bagrow2012}, and in turn comprehend the etiology of epidemics on these systems \cite{Brockmann2009}. In a similar vein, researchers have incorporated the concept of metapopulations \cite{Hanski1998} (a population of populations where in each, mean-field equations suffice to describe the system dynamics) into theoretical epidemiology \cite{Grenfell1997} (for interesting recent examples see \cite{Lund2013} and \cite{Colizza2007}).
 
These similar directions of spatial structure incorporation have naturally been applied to SIS models on networks \cite{Parshani2010, VanMieghem2013, Ferreira2012}, and metapopulations \cite{Arrigoni2002, Ball1999}. From these studies and the current zeitgeist of the field, the current state of the research has implied two salient points: computational power is easily accessible, and the network epidemic modeling is not readily amenable to classical mathematical tools. In this article address these topics, whereby we amalgamate the SIS model with the current impetus toward network/metapopulation modeling through the use of perturbation theory, to quantify the effects of human mobility on an arbitrary network. We show that certain mathematical tools can be brought to bear on epidemic network models yielding accurate analytical approximations to the full temporal dynamics, which are substantiated by the corresponding numerical simulations.

Within the following section we review the analytics of the canonical single-node SIS model, after which we segregate the population into an $M$-node network, whereupon the SIS infection is introduced separately to each. We present a closed-form recursive perturbative solution to the network model; therefrom we calculate explicitly the first-order perturbation leading to our study's main result Eq.~(\ref{eq: first-order-full}). We generalize our result further by analytically solving the short-time dynamics of the network given arbitrary initial conditions, Eq.~(\ref{eq: short_time_approx}), and stitching these to the perturbative solution to yield a full-time approximation to the whole network. Subsequently we compare our result to a test case scenario using real-world population and air traffic data. We then discuss the benefits and limitations of the model and where this work may be applied and built upon.

\section{SIS Model}
In this section we describe the equations which govern the single- and $M$-node models and perform analytical analysis where applicable.

\subsection{Canonical single-node SIS model}
The single-city SIS model considers a large, well mixed, population, of size $N$, in some closed environment where death and birth are neglected. The population is divided into two compartments: susceptible, $S$;  and infected, $I$; where $N= S + I=\textrm{constant}$. Both $S$ and $I$ are discrete variables. Susceptibles may become infected from contact with the infected at a rate $\beta$, and the infected compartment of the population will lose constituents at a rate $\gamma$. The mean-field dynamics of each state is then described by the set of equations:
\begin{equation}\label{eq: Sus_canon}
 \partial_tS(t) = -\frac{\beta}{N}SI + \gamma I,
\end{equation}
\begin{equation}\label{eq: Inf_canon}
 \partial_tI(t) = \frac{\beta}{N}SI - \gamma I,
\end{equation}
where initial conditions are: $S(t=t_0) = S_0>0$, and $I(t=t_0) = I_0>0$. One immediately notes that $\partial_t\left[S(t) + I(t)\right] = 0$, which then ensures that the total population, $N$, is constant for all time. It is noted that in the mean-field equations above, it is implicitly assumed that $S$ and $I$ are treated as continuous variables \cite{Hethcote1989}.

The solution to Eq.~(\ref{eq: Inf_canon}), and in turn Eq.~(\ref{eq: Sus_canon}), as $S(t) = N - I(t)$, is solved in various texts (for example see \cite{Hethcote1989}), so for brevity, the solution is given ($t\geq t_0$):
\begin{equation}\label{eq: Inf_sol}
 I(t) = \frac{I_{\infty}}{1 + V e^{-\chi (t-t_0)}},
\end{equation}
where $\chi = \beta - \gamma$, and $I_{\infty} = \chi N/\beta$ is the stable or endemic state of the infected population; and $V = I_{\infty}/I_0 - 1$ \cite{birthdeathnote}. With respect to Eq.~(\ref{eq: Inf_sol}), for an epidemic to take place (i.e. some finite fraction of the population remains infected in the long-time limit), we require the basic reproductive ratio: $R_0 = \beta/\gamma$ to satisfy $R_0>1$ (which will be assumed henceforth). 

To incorporate a spatial component to the model let us consider an arbitrary network of subpopulations. 

\subsection{Perturbative solution to the $M$-node SIS model}
\begin{figure}[h!]
\centering
\includegraphics[trim = 25mm 15mm 25mm 18mm, clip, scale=0.5]{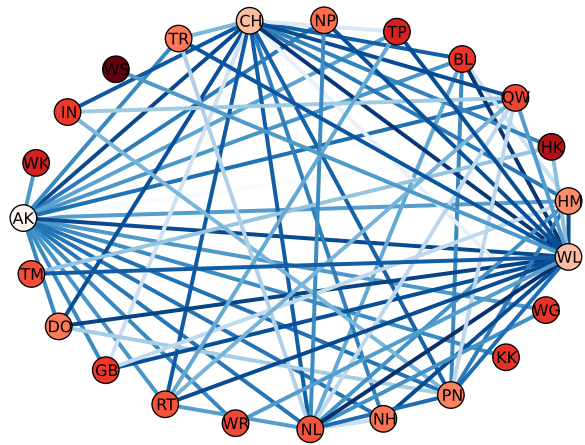} 
\caption{Abstract representation of the New Zealand air traffic network. Each node (23 total) is an airport which services a region, the lighter the shade the more populated the region (reversal of color scale for label clarity, log-scale). Each link/edge on the graph (70 total) represents a flight connection between airports, the darker the link the more transit between those connections (log-scale) \cite{transportnote}. The network has a diameter of three; with the most connected node Auckland (AK, 19 connections), followed by Christchurch (CH, 18 connections), and Wellington (WL, 16 connections). More information on how this network was constructed is contained  in the Appendix, the labels are defined in Tables~\ref{tab: aotearoa_net} and \ref{tab: aotearoa_net_south}.}
\label{fig: nz_network}
\end{figure}

We here further the result found in the previous section through incorporation of an implicit spatial component, by stratifying the large population into $M$ subpopulations (a realistic example is shown Fig.~\ref{fig: nz_network}). Within each subpopulation (which may be regarded as a community, city, or country), the same assumptions stated in the canonical model still hold: the population is sufficiently large, and well mixed. One then allows for mobility between the nodes on the network, where the fraction of persons traveling from city $j$ to $i$ per unit time is given by the transport rate, $\omega_{i\leftarrow j}$.

Upon each node, an SIS virus is introduced, where the $i^{\textrm{th}}$ city has an infectivity rate of $\beta_i$, and recovery rate $\gamma_i$. The mean-field set of equations that then describe the dynamics are given by
\begin{equation}\label{eq: N-city sus}
 \partial_tS_i(t) = -\frac{\beta_i}{N_i}S_iI_i + \gamma_i I_i + \varepsilon \sum_{j =1}^M\left( w_{i\gets j}S_j -w_{j\leftarrow i}S_i\right),
\end{equation}
\begin{equation}\label{eq: N-city inf}
 \partial_tI_i(t) = \frac{\beta_i}{N_i}S_iI_i - \gamma_i I_i + \varepsilon \sum_{j =1}^M\left( w_{i\leftarrow j}I_j -w_{j\leftarrow i}I_i\right).
\end{equation}
where we have defined $\omega_{i\leftarrow i} = 0$. A convenient parameter $\varepsilon(=1)$ is introduced here to keep track of the number of times the perturbation enters below (terms which are linear in $\varepsilon$ are linear in the travel rates etc) \cite{Sakurai}. Summing Eq.~(\ref{eq: N-city sus}) and Eq.~(\ref{eq: N-city inf}), we find that $\partial_tN_i=0$, provided that the total influx and outflux for each node are equal, i.e.,
\begin{equation}\label{eq: net_flux}
\sum_{j=1}^{M}\omega_{j\leftarrow i}N_i=\sum_{j=1}^{M}\omega_{i\leftarrow j}N_j,
\end{equation}
implying the total population $N_i$ of city $i$ is constant for all time. We will assume Eq.~(\ref{eq: net_flux}) to hold henceforth.

To begin the derivation of the perturbative solution, we first assume the influx and outflux of citizens from a given city is small (defined quantitatively later), from there we can define a perturbative solution in terms of the travel rates, namely
\begin{equation}\label{eq: pert-expansion inf}
I_i(t) = \sum^{\infty}_{k=0}\varepsilon^{k}I^{(k)}_i(t) = I_i^{(0)}(t) + \sum^{\infty}_{k = 1}\varepsilon^{k}I^{(k)}_i(t),
\end{equation}
where $I^{(k)}_i(t)$ is the $k^{\textrm{th}}$ order contribution to the perturbative expansion at node $i$; i.e. $I_i^{(1)}$ contains only linear terms in the transport rates, whereas $I_i^{(2)}(t)$ contains only quadratics terms, and so on. In Eq.~(\ref{eq: net_flux}) $I_i^{(0)}(t)$ is given by Eq.~(\ref{eq: Inf_sol}), with replacements $\beta\rightarrow \beta_i$ and $\gamma\rightarrow \gamma_i$ (and therefore $\chi\rightarrow\chi_i$). Explicitly:
\begin{equation}\label{eq: zeroth-order inf sol}
  I_i^{(0)}(t) = \frac{I_{\infty,i}}{1 + V_ie^{-\chi_i (t-t_0)}}. 
\end{equation}
Similarly we can define the perturbative solution to the number of susceptibles in city $i$ as $S_i = \sum^{\infty}_{k=0}\varepsilon^{k}S^{(k)}_i$. Since $N_i = I_i^{(0)}+S_i^{(0)}$, it follows that $I_i^{(k)} = - S_i^{(k)}$ for $k\geq1$. Using this fact and substituting Eq.~(\ref{eq: pert-expansion inf}) into Eq.~(\ref{eq: N-city inf}) and equating factors of $\varepsilon^{k}$, we find for $k=0$, that $\partial_tI^{(0)}_i = \chi_iI^{(0)}_i -\frac{\beta_i}{N_i}\left[I_i^{(0)}\right]^2,$
which is equivalent to Eq.~(\ref{eq: Inf_canon}), and whose solution is therefore given by Eq.~(\ref{eq: zeroth-order inf sol}). For $k\geq1$ we obtain our formal perturbation equations
\begin{eqnarray}\label{eq: kth partial inf}
 \partial_tI_i^{(k)} &-&\left(\chi_i-\frac{2\beta_iI_i^{(0)}}{N_i}\right)I^{(k)}_i = - \frac{\beta_i}{N_i}\sum_{k'=1}^{k-1}I_i^{(k-k')}I_i^{(k')}\nonumber\\
 &+& \sum_{j=1}^{M}\left(\omega_{i\leftarrow j}I_j^{(k-1)}-\omega_{j\leftarrow i}I_i^{(k-1)}\right),
\end{eqnarray}
where we have set $\varepsilon=1$. Thus, we have formally converted the non-linear problem in Eq.~(\ref{eq: N-city sus}) and Eq.~(\ref{eq: N-city inf}) into a set of inhomogeneous, linear equations, Eq.~(\ref{eq: kth partial inf}), with time dependent coefficients. The time dependence of these coefficients in the left-hand-side enters only through the known quantity, $I_i^{(0)}(t)$, whereas the right-hand side depends recursively on the previous perturbation orders.  

With regard to the initial condition of the system, the time $t_0$, viz. Eq.~(\ref{eq: zeroth-order inf sol}), need not be the true initial time, but rather some time at which $\vec{I}(t)(\equiv[I_1(t), I_2(t),\cdots, I_M(t)]^T)$ is known. We will, in a subsequent section, utilize this freedom of choice in $t_0$ to improve upon the results in this section.

We proceed by expressing a formal solution to Eq.~(\ref{eq: kth partial inf}) through employment of the integrating factor method. Firstly, we define the so-called integrating factor: $\exp(B_i(t))$, where $B_i(t) = - \int^t_{t_0}\left[\chi_i - 2\beta_iI_i^{(0)}(t')/N_i\right]dt'.$ Then the formal solution to the $k^{\textrm{th}}$-order perturbative term is $I^{(k)}_i(t) = \exp\left(-B_i(t)\right)\left[\int^t_{t_0}\exp\left(B_i(t')\right) g_i^{(k-1)}(t')dt' +G_i^{(k)}\right],$ where, from the initial conditions: $I_i^{(k)}(t=t_0) = 0$, we have $G_i^{(k)} = 0$. The function $g_i^{(k-1)}(t)$ is defined as
\begin{eqnarray}\label{eq: little-g formal}
g_i^{(k-1)}(t) &=&  - \frac{\beta_i}{N_i}\sum_{k'=1}^{k-1}I_i^{(k-k')}I_i^{(k')}\nonumber\\
 &+& \sum_{j=1}^M\left(\omega_{i\leftarrow j}I_j^{(k-1)}-\omega_{j\leftarrow i}I_i^{(k-1)}\right).
\end{eqnarray}
Interestingly, we are able to calculate $B_i(t)$ explicitly. Using Eq.~(\ref{eq: zeroth-order inf sol}), we can write $B_i(t) = -\chi_i(t-t_0)+(2\beta_iI_{\infty,i})/(N_i)\int^{t}_{t_0}(1+V_ie^{-\chi_i(t'-t_0)})^{-1}dt'.$
We solve this to yield the solution $B_i(t) = \ln\left[e^{\chi_i (t-t_0)}\left(\frac{1+V_ie^{-\chi_i (t-t_0)}}{1+V_i}\right)^2\right].$ Using the solution for $B_i(t)$ and Eq.~(\ref{eq: little-g formal}) and substituting this into the formal solution given, we explicitly obtain the $k^{\textrm{th}}$ order perturbation, namely 
\begin{eqnarray}\label{eq: formal kth inf full}
 I^{(k)}_i(t) &=& e^{-\chi_i(t-t_0)}\left(1+V_ie^{-\chi_i (t-t_0)}\right)^{-2}\bigg[\int^t_{t_0} e^{\chi_i(t'-t_0)}\nonumber\\
&&\times\left(1+V_ie^{-\chi_i (t'-t_0)}\right)^{2}g_i^{(k-1)}(t')dt'\bigg].
\end{eqnarray}
With this closed form expression, we are able to calculate any order perturbation we require, recursively. Namely, starting from the zeroth-order solution, Eq.~(\ref{eq: zeroth-order inf sol}), we can insert this into Eq.~(\ref{eq: little-g formal}), the result of which is then input into Eq.~(\ref{eq: formal kth inf full}) to find the first-order perturbation (shown explicitly in the following section). To find the next order, one uses the first-order result in place of the zeroth-order solution to, following the outlined algorithm, arrive at the second-order perturbation. This operation may be repeated until the desired number of orders are achieved. Then the orders are summed, viz. Eq.~(\ref{eq: pert-expansion inf}) (with $\varepsilon=1$), to gain the final solution to the infected population contained in city $i$.

\subsection{Explicit first-order perturbation}
Let us now calculate the first order perturbation term $k=1$. Then the function $g^{(0)}_i$, see Eq.~(\ref{eq: little-g formal}), is explicitly: $g^{(0)}_i(t)=\sum_j\left(\omega_{i\leftarrow j}I^{(0)}_j-\omega_{j\leftarrow i}I^{(0)}_i\right),$ such that Eq.~(\ref{eq: formal kth inf full}), using Eq.~(\ref{eq: zeroth-order inf sol}), becomes 
\begin{equation}\label{eq: I-with-Q}
I^{(1)}_i = \frac{e^{-\chi_i(t-t_0)}}{\left(1+V_ie^{-\chi_i(t-t_0)}\right)^2}\sum_{j=1}^M\left[\omega_{i\leftarrow j}Q_{ij}-\omega_{j\leftarrow i}Q_{ii}\right],
\end{equation}
where
\begin{equation}\label{eq: q-integral}
Q_{ij} = I_{\infty,j}\int^{t}_{t_0} e^{\chi_i (t'-t_0)}\frac{\left(1+V_ie^{-\chi_i(t'-t_0)}\right)^2}{1+V_je^{-\chi_j(t'-t_0)}}dt'.
\end{equation}
The quantity $Q_{ij}$ may be expressed in terms of hypergeometric functions \cite{Abramowitz1972}. 

For the scope of this manuscript, let us analyze the case where we shall assume that all infection and recovery rate parameters are independent of the city, namely $\beta_i=\beta_j=\beta$ and $\gamma_i=\gamma_j=\gamma$. Explicitly evaluating Eq.~(\ref{eq: q-integral}), we find that
\begin{eqnarray}
Q_{ij} = \frac{-I_{\infty,j}}{\chi}\Bigg[&&\frac{\left(V_i-V_j\right)^2}{V_j}\ln\left(\frac{1+V_je^{-\chi (t-t_0)}}{1+V_j}\right)+1\nonumber\\
&&-\chi (t-t_0)(2V_i-V_j)-e^{\chi (t-t_0)}\Bigg].
\end{eqnarray}
This leads to the explicit first-order perturbative contribution to Eq.~(\ref{eq: pert-expansion inf}),
\begin{eqnarray}\label{eq: first_pert_flux}
I^{(1)}_{i} &=& \frac{e^{-\chi (t-t_0)}}{\chi\left(1+V_ie^{-\chi (t-t_0)}\right)^2}\Bigg(\sum_{j=1}^M\omega_{i\gets j}I_{\infty,j}\times\nonumber\\
&&\bigg[\chi(2V_i-V_j)(t-t_0)-\nonumber\\
&&\frac{(V_i-V_j)^2}{V_j}\ln\left(\frac{1+V_je^{-\chi(t-t_0)}}{1+V_j}\right)\bigg]\nonumber\\
&&-\sum_{j=1}^M\omega_{j\gets i}I_{\infty,i}\chi V_i(t-t_0)\Bigg),
\end{eqnarray}
where we have used Eq.~(\ref{eq: net_flux}). Eqs.~(\ref{eq: zeroth-order inf sol}) and (\ref{eq: first_pert_flux}), with $I_i(t) = I_i^{(0)}(t)+I_i^{(1)}(t)$ constitute the first-order solution to the SIS epidemic.

If instead of zero net nodal flux, Eq.~(\ref{eq: net_flux}), we instate the more restrictive clause of detailed balance \cite{Brockmann2009}, $\omega_{i\gets j}N_j=\omega_{j\gets i}N_i$, we have that $\omega_{i\leftarrow j}I_{\infty,j} = \omega_{j\leftarrow i}I_{\infty,i}$. Using this relation, we can write out the first order perturbation, Eq.~(\ref{eq: first_pert_flux}), explicitly as
\begin{eqnarray}\label{eq: first_order_explicit}
I^{(1)}_{i}&=&\frac{I_{\infty,i}e^{-\chi (t-t_0)}}{\chi\left(1+V_ie^{-\chi (t-t_0)}\right)^2}\sum_{j=1}^M\omega_{j\leftarrow i}(V_i - V_j)\times\nonumber\\
&&\left[\chi(t-t_0) -\frac{V_i-V_j}{V_j}\ln\left(\frac{1+V_je^{-\chi (t-t_0)}}{1+V_j}\right)\right].
\end{eqnarray}
Summing this with the zeroth-order solution we reap the first-order perturbative approximation to the number of infected in city $i$ at time $t$:
\begin{widetext}
\begin{equation}\label{eq: first-order-full}
 I_i(t) = \frac{I_{\infty,i}}{1 + V_ie^{-\chi (t-t_0)}}\Bigg(1 + \frac{e^{-\chi (t-t_0)}}{\left(1+V_ie^{-\chi (t-t_0)}\right)}\sum_{j=1}^M\frac{\omega_{j\leftarrow i}}{\chi}(V_i - V_j)\left[\chi (t-t_0) -\frac{V_i-V_j}{V_j}\ln\left(\frac{1+V_je^{-\chi (t-t_0)}}{1+V_j}\right)\right]\Bigg) + \mathcal{O}(w_{i\leftarrow j}^2),
\end{equation}
\end{widetext}
where 
\[
  V_i = \frac{I_{\infty,i}}{I_{0,i}}-1,
\]
and 
\[
 I_{\infty,i} = \frac{N_i\chi}{\beta}.
\]
The first-order approximation to $I_i(t)$, namely, Eq.~(\ref{eq: first-order-full}), is valid when the perturbation due to mobility is ``small''. To quantify explicitly the validity of Eq.~(\ref{eq: first-order-full}) we introduce the approximate validity indicator:
\begin{equation}\label{eq: valid-ratio}
 C_i(t_0) = \sum_{j=1}^M\frac{\omega_{j\leftarrow i}}{\chi}\left\lvert V_i - V_j\right\rvert = \sum_{j=1}^M\frac{\omega_{j\leftarrow i}}{\beta}\left\lvert\frac{N_i}{I_{0,i}} - \frac{N_j}{I_{0,j}}\right\rvert.
\end{equation}
For a given system, when $C_i(t_0)\approx 0$ the zeroth-order solution is valid for $I_i(t)$; when $C_i(t_0)\sim \mathcal{O}(1)$ the first-order solution is an accurate approximation to Eq.~(\ref{eq: N-city inf}). For $C_i(t_0)\gg1$, Eq.~(\ref{eq: first-order-full}) breaks down. It should be noted that $C_i(t_0)$ may be valid for node $i$, but the corresponding indicator for some other node $j$ may not be. In this case, Eq.~(\ref{eq: first-order-full}) would be reasonable still for $I_i(t)$, but not for $I_j(t)$, therefore caution is advised. In particular we point out that, besides the travel rate $\omega_{j\leftarrow i}$ (in units of $\beta$), also the fraction of initially infected for each node enters Eq.~(\ref{eq: valid-ratio}) in a non-trivial way. Eq.~(\ref{eq: valid-ratio}) requires that every neighboring node have a finite fraction of initially infected for Eq.~(\ref{eq: first-order-full}) to be valid. This stems from the fact that if a node is initially uninfected, the transport of infected persons into that node, see Eq.~(\ref{eq: N-city inf}), is no longer small (compared to the infectivity and recovery term), i.e. the base assumption of our perturbative approach is violated, therefore Eq.~(\ref{eq: first-order-full}) breaks down. To remedy this initial condition restriction, we turn to linearization of the short-time dynamics in the next subsection.

\subsection{Short-time approximation}
We utilize the prerogative in the choice of $t_0$ in the previous section in order to generalize Eq.~(\ref{eq: first-order-full}). We do this via approximating the short-time regime to allow for any initial state of the system, not only that all nodes be infected as required by the first-order perturbation result. We begin by neglecting the quadratic term $I_i^2N_i^{-1}$ in Eq.~(\ref{eq: N-city inf}) (as it is generally small for short times compared to the linear term), thereby linearizing it to 
\begin{equation}
 \partial_tI_i(t) \approx \bigg(\chi -\sum_{j =1}^M \omega_{j\leftarrow i}\bigg) I_i + \sum_{j =1}^M \omega_{i\leftarrow j}I_j.
\end{equation} 
This can be written as $\partial_t\vec{I}(t)\approx(\boldsymbol{\Omega} + \boldsymbol{\chi})\vec{I}(t)$, where $\Omega_{ij} = \omega_{i\gets j}$ (for $i\neq j$), and $\Omega_{ii}= -\sum_j \omega_{j\gets i}$. For the second matrix, $\boldsymbol{\chi} = \chi\boldsymbol{I}$ ($\boldsymbol{I}$ is the identity matrix). Through the Baker-Campbell-Hausdorff formula, and the commutativity of $\boldsymbol{\chi}$ and $\boldsymbol{\Omega}$, the general solution to this set of equations is
\begin{equation}\label{eq: short_time_approx}
\vec{I}(t) =\exp(\boldsymbol{\Omega}t)e^{\chi t}\vec{F}_0,
\end{equation}
where $\vec{F}_0 = \vec{I}(t=0)$ is the actual initial condition \cite{RK_note}. Note that $\vec{F}_0$ is different to the initial conditions used for the first-order perturbation result, i.e. $\vec{I}_0 = \vec{I}(t=t_0)$. In Eq.~(\ref{eq: short_time_approx}) the zeroth-order dynamics are captured via $\exp(\chi t)\vec{F}_0$, and the correction factor to the zeroth-order, captured in $\exp(\boldsymbol{\Omega}t)$, due to travel. 

To find a limiting time for which Eq.~(\ref{eq: short_time_approx}) is valid, consider the following: we can recast Eq.~(\ref{eq: short_time_approx}) as 
\begin{equation}\label{eq: short_time_eigenmode}
\vec{I}(t) = e^{\chi t}\sum_\alpha \vec{r}_\alpha \exp(\lambda_\alpha t)\vec{\ell}_\alpha\cdot\vec{F}_0,
\end{equation}
where the subscript $\alpha$ labels the eigenmode of the eigenvalue ($\lambda_\alpha$), to the corresponding left ($\vec{\ell}_\alpha$) and right ($\vec{r}_\alpha$) eigenvectors of $\boldsymbol{\Omega}$ \cite{BCH_note}. We can approximate these exact linear dynamics by considering the contribution of only the leading eigenmode, $\alpha_0 = 0$ with $\lambda_0=0$ \cite{eigen_bounds}. Then we have $\vec{\ell}_0 = [1,1,\cdots,1]^T$, and $\vec{r}_0 = N_{\textrm{tot}}^{-1}[N_1,N_2,\cdots,N_M]^T$, where $N_{\textrm{tot}} = \sum_{j=1}^MN_j$. So, per node, Eq.~(\ref{eq: short_time_eigenmode}) simplifies to $I_i(t)\approx N_{\textrm{tot}}^{-1}N_iF_0^{\textrm{tot}}\exp(\chi t)$, where $F_0^{\textrm{tot}} = \sum_{j=1}^MF_{0,j}$. We wish to investigate where the linear approximation is valid, by comparing the linear and quadratic terms: $\beta/N_i I_i^2 \ll \chi I_i,$ or $I_i \ll N_i \chi\beta^{-1}\equiv I_{\infty,i}$. Using the $\alpha=0$ linear approximation for $I_i$ (as described above), we get that the linear approximation should be valid for $\exp(\chi t) N_iN_{\textrm{tot}}^{-1} F_0^{\textrm{tot}} \ll N_i \chi\beta^{-1}$ where the $N_i$ drops out, leaving us with $\exp(\chi t) \ll (\chi\beta^{-1}) / (F_0^{\textrm{tot}}/N_{\textrm{tot}})$
which we recognize as the ratio between the asymptotic fraction of infecteds to the initial one. Thus we expect the linear approximation to break down as this
limit saturates, i.e. at 
\begin{equation}\label{eq: short_time_limit}
t_s = \chi^{-1} \ln[(\chi/\beta) / (F_0^{\textrm{tot}}/N_{\textrm{tot}})]. 
\end{equation}
A positive $t_s$ indicates that we need the initial linear stage, before switching to the non-linear dynamics in the perturbative approximation. A negative $t_s$, on the other hand, indicates that we can skip the linear step, and go directly to the nonlinear stage. Henceforth, $t_s$ will be referred to as the short-time limit.

If we are to "stitch" Eq.~(\ref{eq: short_time_approx}) to Eq.~(\ref{eq: first-order-full}), we require that Eq.~(\ref{eq: first-order-full}), be valid, i.e. $C_i(t_0)\sim \mathcal{O}(1)$, where $t_0$ shall now be known as the stitch time. This criterion sets a soft lower bound to the use of Eq.~(\ref{eq: short_time_approx}). We establish a cut-off value of the validity indicator to be $C_{\textrm{cut},i} = C_i(t=t_c)\sim \mathcal{O}(1)$. Then the time-frame for a suitable stitching is $t_c<t_0<t_s$. This stitched approximation allows us to model the full temporal dynamics of the network irrespective of the initial conditions.

To illustrate Eq.~(\ref{eq: short_time_approx}) and Eq.~(\ref{eq: short_time_eigenmode}), we construct a simple system of two nodes, whose the eigenvalues are $\boldsymbol{\Omega}$ are $\{\lambda_1, \lambda_2\} = \{0, -(\omega_{1\gets 2}+\omega_{2\gets 1})\}$ (see \cite{eigen_bounds}), with the respective eigenvectors: $\vec{v}_1 = [\omega_{1\gets 2}\omega_{2\gets 1}^{-1}, 1]^{T}$ and $\vec{v}_2 = [-1, 1]^{T}$. If we set the initial conditions, $\vec{I}(t=0) = [F_{0,1}, 0]^{T}$, then the nodal short-time evolution is: $I_1(t) = F_{0,1}\omega_{T}^{-1}[\omega_{2\gets 1}\exp(\chi t)+\omega_{1\gets 2}\exp([\chi-\omega_T]t)]$, $I_2(t) = F_{0,1}\omega_{2\gets 1}\omega_{T}^{-1}[\exp(\chi t)-\exp([\chi-\omega_T]t)]$; where $\omega_T = \omega_{2\gets 1} + \omega_{1\gets 2}$.

\section{SIS Epidemic in New Zealand (\emph{Aotearoa})}

We proceed by measuring the performance of our analytical expressions, Eq.~(\ref{eq: first-order-full}) and Eq.~(\ref{eq: short_time_approx}), with realistic population \cite{NZstats} and air transport data \cite{airtraf_note} through a test case scenario, an SIS epidemic in New Zealand (NZ). Firstly, we suppose an infectious virus ($\beta = 0.15$ $\textrm{day}^{-1}$, $\gamma = 0.10$ $\textrm{day}^{-1}$, $R_0 = 1.5$) has introduced itself within the Auckland region (see the Appendix for population statistics) and assume people (and therefore the virus), both susceptible and infected, move between nodes via the air traffic transport network where there is no transport internationally with NZ (which one could think of as a quarantine measure \cite{chatham}). We consider two networks to illustrate the work done herein, firstly, a simplified model where we recompose NZ into two subpopulations, Auckland, and the remainder of NZ. In the second scenario we consider the full NZ network (as shown in Fig. \ref{fig: nz_network} and described in the Appendix).

\subsection{NZ two-node system}
\begin{figure}[b]
\includegraphics[scale=0.7]{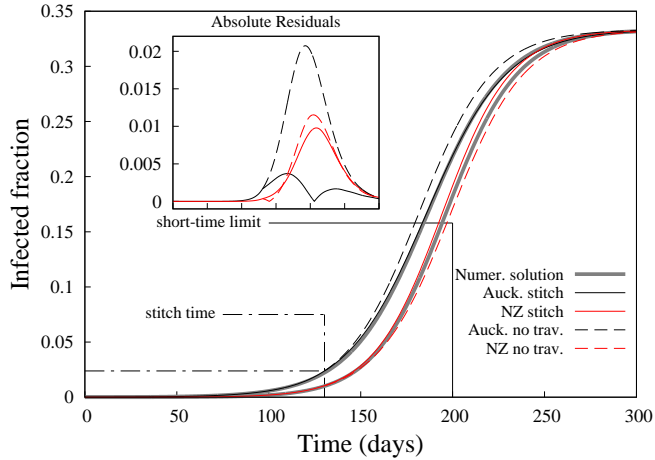}
\caption{\label{fig: two_node} The infected fraction of the populace over time (due to an SIS epidemic) for each node for an internationally quarantined New Zealand, apportioned into two subpopulations: Auckland, and the remainder of the country. One clearly notes that in both populations, the stitched first-order perturbation, Eq.~(\ref{eq: first-order-full}) and Eq.~(\ref{eq: short_time_approx}) (solid red/black line), approximates the numerical solution well (solid gray line), compared to the stitched zeroth-order solution, Eq.~(\ref{eq: zeroth-order inf sol}) and Eq.~(\ref{eq: short_time_approx}) (dashed red/black line). INSET: Absolute residuals of the analytical calculations, with respect to the numerical solution. For further explanation and auxiliary parameters see the main text.}
\end{figure}

In the event of an outbreak given the virus parameters mentioned above, we seek to understand how the country's largest city, Auckland (\emph{T\={a}maki Makaurau}, subscript ``Auck''), is affected and/or affects the remainder of NZ (subscript ``rem''). We divide NZ into these two subpopulations (thereby assuming that each of these populations is well mixed). The documented transport rates between these two nodes are $\omega_{\textrm{rem}\leftarrow \textrm{Auck}} \approx 5.5\times10^{-3}$ $\textrm{day}^{-1}$, and $\omega_{\textrm{Auck}\leftarrow \textrm{rem}} \approx 3.2\times10^{-3}$ $\textrm{day}^{-1}$ \cite{transportnote}, which are indeed small -- as required by our perturbation assumption. We set the initial infected population to $F_{0,\textrm{Auck}}=100$, and $F_{0,\textrm{rem}}=0$, for Auckland and the remainder of NZ, respectively. Given these initial conditions and virus parameters, the short-time limit is $t_s\approx200$ days. We use Eq.~(\ref{eq: short_time_approx}) to model the network until all nodes in the network are infected ($I_i(t)\geq1,$ $\forall i$), and the validity indicator for Auckland, $C_{\textrm{Auck}}\leq 2$. At this time ($t_0\approx108$ days) we use the state of the system as the initial condition for Eq.~(\ref{eq: first-order-full}), and model the remaining infection dynamics for both nodes. We also compare our stitched first-order result to the stitched zeroth-order result (using Eq.~(\ref{eq: zeroth-order inf sol}) and Eq.~(\ref{eq: short_time_approx})). This scenario of "no travel" can be likened to a situation of quarantine when all nodes have been found to be infected. It is also a measure of how effective the zeroth-order solution, Eq.~(\ref{eq: zeroth-order inf sol}), is at estimating the time-evolution of the epidemic. The stitched first-order and the stitched zeroth-order perturbation approximations are compared to the numerical solution, $\vec{I}_{\textrm{num}}(t)$ (calculated by the Runge-Kutta 4$^{\textrm{th}}$ order algorithm) in Fig. \ref{fig: two_node}.
We note that, in Fig. \ref{fig: two_node}, our analytical result conforms well to the numerical result, where the absolute residuals ($|\Delta_j(t)|$, see Appendix, Eq.~(\ref{eq: residues})) between the first-order correction and the zeroth-order are given in the inset. The zeroth-order solution performs poorly at estimating the interim dynamics of the epidemic, especially underestimating the fraction of infected in Auckland.

\subsection{Full NZ network}
\begin{figure}[b]
\includegraphics[scale=0.7]{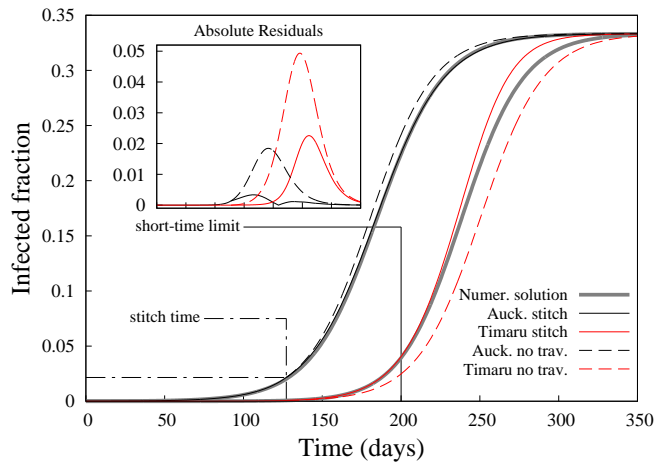}
\caption{\label{fig: aotearoa_nodes} Infected fraction over time for the Auckland and Timaru regions of an internationally quarantined NZ - illustrated in Fig. \ref{fig: nz_network}. These data demonstrate the capabilities of the stitched first-order perturbation, Eq.~(\ref{eq: first-order-full}) and Eq.~(\ref{eq: short_time_approx}) (solid red/black line), on a realistic network (23 airport nodes of NZ, see Appendix) compared to the numerical solution (solid gray line). For further explanation and auxiliary parameters see the main text.}
\end{figure}

To assess further the scope of our perturbation result, we reconstruct the system to include all available airports, giving a total of 23 nodes (see Fig. \ref{fig: nz_network} and the Appendix). For this system we assume the same virus as before, with the same initially infected, whom have been introduced initially to the Auckland node (this again gives a short-time limit of $t_s\approx200$ days). Again, we use Eq.~(\ref{eq: short_time_approx}) to simulate the short-time regime until all nodes are infected ($I_i(t)\geq1$, $\forall i$), and the validity indicator for Auckland $C_{\textrm{Auck}}\leq2$. At this time, $t_0\approx127$ days, we use Eq.~(\ref{eq: first-order-full}) to model the dynamics given the current state of the system. We plot these results for Auckland compared to the numerical solution and the stitched zeroth order approximation, Eqs.~(\ref{eq: zeroth-order inf sol}) and (\ref{eq: short_time_approx}), in Fig. \ref{fig: aotearoa_nodes}.

The choice of $t_0$ here favors a good validity indicator for Auckland. To understand the repercussion of this on other nodes in the network, alongside the infection dynamics of Auckland in Fig. \ref{fig: aotearoa_nodes}, we have plotted the dynamics of Timaru (\emph{Te Tihi-o-Maru}). Timaru has the highest validity at the stitch time, $C_{\textrm{Tim}}\approx6.39$. This larger validity indicator makes for a worse approximation, as shown in Fig.~(\ref{fig: aotearoa_nodes}), compared to Auckland with respect to the numerical solution. On the other hand, the stitched zeroth-order fares more poorly given this $t_0$. This highlights the subtle consequences in the choice of $t_0$.

\section{Discussion and Conclusion}
Within this manuscript, we have derived a closed form, recursive, perturbative solution to an SIS epidemic on an arbitrary network, stated in Eqs. (\ref{eq: little-g formal}) and (\ref{eq: formal kth inf full}). We have proceeded to explicitly calculate the first-order perturbation to the population of infected persons in the $i^{\textrm{th}}$ city as a function of time, to wit, Eq.~(\ref{eq: first-order-full}), and then provided a quantitative benchmark under what conditions this solution is accurate: the validity indicator, Eq.~(\ref{eq: valid-ratio}). 

We have generalized Eq.~(\ref{eq: first-order-full}), to include the arbitrary initial condition of the network through the linearization of the short-time dynamics, Eq.~(\ref{eq: short_time_approx}). We quantified a short-time limit, $t_s$ Eq.~(\ref{eq: short_time_limit}), for which Eq.~(\ref{eq: short_time_approx}) is valid. The results of Eq.~(\ref{eq: short_time_approx}) are stitched to Eq.~(\ref{eq: first-order-full}), at the stitch time, $t_0(<t_s)$ (see main text for further remarks). This gives an approximation to the full network dynamics irrespective of initial conditions.  

To verify our derived results we simulated an SIS epidemic on an internationally quarantined New Zealand (see Fig. \ref{fig: nz_network}). This comparison served a two-fold objective; firstly the use of documented air traffic data \cite{airtrafficstats} showed that in this medium of transport, our base assumption: transport rates between communities are small, is indeed reasonable and accurate as a perturbation parameter. Secondly, it serves to show the extent of use of our stitched first-order result; namely that it performs well on a realistic, moderately sized (23 nodes) network.

The derivation makes no assumptions on the type of the network, whether it be a real-world network, regular lattice, a random E-R graph, or a scale-free network \cite{Newman2003}. It has also only assumed, besides the detailed balance condition (see Eq.~(\ref{eq: first_pert_flux}) for an expression without this assumption), that the population of each node is large enough such that the mean-field nature of Eq.~(\ref{eq: Inf_sol}) is true. Therefore the nodes may be seen as communities or countries, rather than only cities. This generality is advantageous for future investigations of SIS epidemics on complex networks/metapopulations as this work may be used in parallel as a confidence measure. But caution is advised. The success of the stitch approximation on this network is due to the magnitude of the travel rates and the diameter of the network. The NZ network has a low diameter of three, with reasonable travel rates, so in turn the short-time approximation is able to "seed" every node, enough so that $C_{\textrm{Auck}}(t_0)\sim \mathcal{O}(1)$ while $t_0<t_s$. One could envisage a sparse network with a large diameter and slow travel rates; such that $C_{i}(t_0)\sim \mathcal{O}(1)$ for $t_s<t_0$. This could then lead to the breakdown of our calculations. So therefore it would be of interest to understand how higher order perturbation terms may regularize our first-order approximation.

One of the striking benefits of the solution is that one has an analytical expression for $I_i(t)$ at all times and as such naturally out-performs usual investigative methods of numerical integration. In this way this solution can be used to gauge parameter sets of large numerical simulations.

These calculations were built upon the assumption of large populations, where mean-field approximations are valid. A natural extension would be the effect of stochasticity for low populations. This may be found through an analytic perturbative solution of the associated master equation akin to that defined for an SIR epidemic in the work of Hufnagel et al. \cite{Hufnagel2004}.

Although the analytics have been developed under the guise of epidemic modeling, this mathematical frame-work may be conveniently adopted by other interdisciplinary fields with population growth and metapopulation structure, for example Theoretical Ecology and the concept of island colonization \cite{Levins1969}.

In conclusion, we hope the mathematical framework determined herein will shift part of the academic interest of epidemics on networks from large scale numerical simulations back to the bedrock of analytical analysis.

\begin{acknowledgements}
The authors would like to thank Erik Lagerstedt, Olivia Woolley-Meza, Sigur\dh ur \AE. J\'onsson, and for stimulating discussions and the reviewers for constructive criticism regarding the work herein.
\end{acknowledgements}
\appendix

\section{Appendix}
Within the Appendix we offer more statistics on our analysis of the epidemic on the full NZ network, and instructions on how the network was constructed from census and air traffic data.
\subsection{NZ Network Statistics}
\begin{figure}[b]
\includegraphics[scale=0.72]{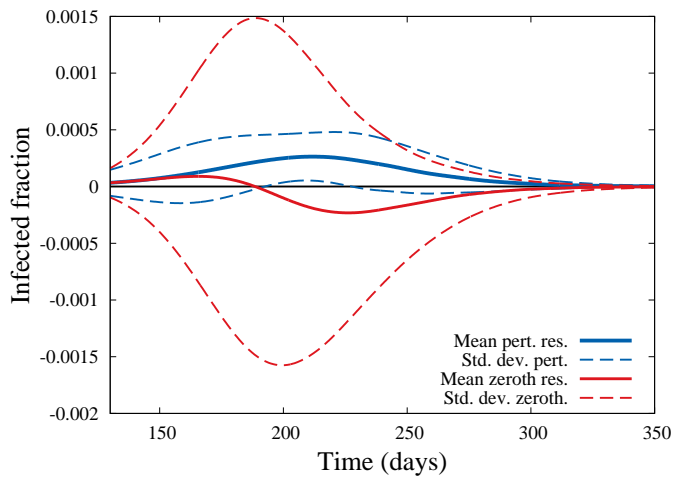}
\caption{\label{fig: aotearoa_stats} Illustrated above are the residue statistics as a measure of performance of the first and zeroth order approximations in comparison to the numerical solution (see also Fig. \ref{fig: aotearoa_nodes} for more information). Plotted is $\bar{\Delta}(t)\pm \sigma_\Delta(t)$ of the NZ network (Tables \ref{tab: aotearoa_net} and \ref{tab: aotearoa_net_south}), given a virus $R_0 = 1.5$, $\beta = 0.15$  $\textrm{day}^{-1}$, initially introduced to Auckland, $F_{\textrm{Auck}}(0)=100$. The measurement is conducted at the time of stitching, the time at which the epidemic has spread to all nodes ($I_i(t)>1$, $\forall i$), and the validity indicator, Eq.~(\ref{eq: valid-ratio}), for Auckland is $C_{\textrm{Auck}}\leq2$. The ordinate axis shows the fraction of the total population.}
\end{figure}

In the full system Eqs.~(\ref{eq: first-order-full}) and (\ref{eq: short_time_approx}) model the dynamics well for Auckland and Timaru, in comparison to Eqs.~(\ref{eq: zeroth-order inf sol}) and (\ref{eq: short_time_approx}) (as shown in Fig. \ref{fig: aotearoa_nodes}), but we seek the performance over the entire network. Due to the size of the system, we have opted to show the residue statistics as a measure of performance over all nodes. We define the residue of node $j$ as 
\begin{equation}\label{eq: residues}
\Delta_j(t)=I_{j}(t)-I_{\textrm{num},j}(t),
\end{equation}
where $I_{j}(t)$ is either the first-order solution or the zeroth-order. The mean residue is then $\bar{\Delta}(t) = M^{-1}\sum_{j=1}^M\Delta_j(t)$; and the standard deviation is $\sigma_\Delta(t) = \sqrt{M^{-1}\sum_j(\Delta_j - \bar{\Delta})^2}$. We see holistically in Fig. \ref{fig: aotearoa_stats} that the first-order performs better than the zeroth-order solution, on average, for this system.

\subsection{NZ Network Construction and Parameters}
For the case study within this manuscript, NZ has been recomposed into 23 nodes, each identified by the airport which services the node. Each node is constructed through a combination of the census data \cite{NZstats} and air traffic data \cite{airtrafficstats}. The explicit population of each node is shown in Tables \ref{tab: aotearoa_net} and \ref{tab: aotearoa_net_south}. The node is composed of census defined "regions" (numbering 66), and how each census region is billeted to which airport is defined through the following criteria: 1) if a region has only one airport, that airport services the region; 2) If a region has two or more airports, the total influx/outflux of those airports are combined to make one airport node (e.g. Kaitai-Kerikeri); 3) if a region has no airports, and several adjacent airports, the largest airport services that region; 4) all other situations are made on a case by case basis and are highlight via '*' in Tables \ref{tab: aotearoa_net} and \ref{tab: aotearoa_net_south}. The travel rates between nodes are described in Ref. \cite{transportnote}.
\begin{table*}[b]
\caption{\label{tab: aotearoa_net}%
North Island network of New Zealand. First column defines which airport (with designated label used in Fig. \ref{fig: nz_network}) is used by which regions, given in the second column. The third column gives the population for each region, and the fourth is the total population using the corresponding airport.}
\begin{ruledtabular}
\begin{tabular}{ccccc}
Node Airport&Regions of Node&Region Pop.&Node Pop.\\
\hline
Kaitai-Kerikeri (KK)&Far North&58500&58500\\
\hline
Whangarei (WR)&Whangarei& 80500& 80500\\
\hline
Auckland (AK)&Kaipara& 119150&  \\
& Auckland city & 1486000 &\\
& Hauraki & 18750 &\\
& Thames-Coromandal* & 27000 &\\
& Waikato & 64300 & 1615200\\
\hline
Hamilton (HM)&Hamilton city&145600&  \\
& Matamata-Piako & 32000 &\\
& Waipa & 46100 &223700\\
\hline
Tauranga (TR)&Tauranga city&115700&  \\
&Western BOP&45800& 161500\\
\hline
Rotorua (RT)&Rotorua&68900&  \\
&South Waikato&22900& 91800 \\
\hline
Whakatane (WK)&Whakatane&34500&\\
&Kawerau&6940&41440\\
\hline
Gisborne (GB)&Gisborne&46600&\\
&Opotiki&8950&\\
&Wairoa&8350&63900\\
\hline
Taupo (TP)&Taupo&34100&\\
&Otorohanga&9320&43420\\
\hline
New Plymouth (NP)&New Plymouth&73800&\\
&Waitomo&9630&\\
&Ruapehu&13400&\\
&Stratford&9170&\\
&Sth. Taranaki&26900&132900\\
\hline
Napier-Hastings (NH)&Napier city&57800&\\
&Hastings&75500&\\
&Centrl Hawkesbay*&13500&146800\\
\hline
Whanganui (WG)&Whanganui&43500&\\
&Rangitikei&14800&58300\\
\hline
Palmerston Nrth. (PN)&Palmerston Nrth. city& 82100&\\
&Manawatu&30000&\\
&Tararua&17700&\\
&Horowhenua&30700&\\
&Masterton*&23500&184000\\
\hline
Wellington (WL)&Wellington city&200100&\\
&Kapiti Coast&49800&\\
&Porirua city*&52700&\\
&Upperhutt city& 41500&\\
&Lowerhutt city&103000&\\
&Sth. Wairarapa&9420&\\
&Carterton&7650&464170\\
\end{tabular}
\end{ruledtabular}
\end{table*}

\begin{table*}[b]
\caption{\label{tab: aotearoa_net_south}%
South Island network of New Zealand. See Table \ref{tab: aotearoa_net} for discussion of columns.}
\begin{ruledtabular}
\begin{tabular}{ccccc}
Node Airport&Regions of Node&Region Pop.&Node Pop.\\
\hline
Nelson (NL)&Nelson city&46200&\\
&Tasman&48100&94300\\
\hline
Westport (WS)&Buller& 10100&10100\\
\hline
Hokitika (HK)&Westland&8960&  \\
& Grey& 13900&22860\\
\hline
Christchurch (CH)&Christchurch city&367700&\\
&Waimakariri&48600&\\
&Selwyn&41100&457400\\
\hline
Timaru (TM)&Timaru&44700&\\
&Ashburton&30100&\\
&Mackenzie&4050&\\
&Waimate&7630&86480\\
\hline
Queenstown-Wanaka (QW)&Queenstown lakes& 28700&\\
&Central Otago&18400&\\
&Southland&29600&76700\\
\hline
Blenheim (BL)&Marlborough&45600&\\
&Kaikoura&3850&\\
&Hurunui&11300&60750\\
\hline
Dunedin-Oamaru (DO)&Waitaki&20900&\\
&Dunedin city&126000&\\
&Clutha&17550&164450\\
\hline
Invercagill (IN)&Invercagill&53000\\
&Gore*&12300&65300\\
\end{tabular}
\end{ruledtabular}
\end{table*}

\bibliographystyle{apsrev4-1}%
\bibliography{sis-refs.bib}

\begin{thebibliography}{10}
\bibitem{BCH_note}%
  \BibitemOpen
  \bibinfo {note} {For the case of different infectivity and recovery rates for
  the different cities, the matrix $\mathbf{\chi}$ is replaced by a diagonal
  matrix with elements $\chi_i$. For this case $\mathbf{\chi}$ and
  $\mathbf{\Omega}$ do not commute in general, but the Baker-Campbell-Hausdorff
  formula still applies. This formula then requires us to evaluate a set of
  commutation relations in order to get an explicit expression for
  $\vec{I}(t)$.}%
  \bibAnnoteFile{Stop}{BCH_note}%
\bibitem{airtraf_note}%
  \BibitemOpen
  \bibinfo {note} {The transport rates we use \cite{airtrafficstats} satisfy
  the detailed balance condition, as required for the validity of Eq.~(\ref{eq:
  first-order-full}).}%
  \bibAnnoteFile{Stop}{airtraf_note}%
\bibitem{chatham}%
  \BibitemOpen
  \bibinfo {note} {We consider travel to and from the Chatham islands to be
  international travel and is therefore omitted.}%
  \bibAnnoteFile{Stop}{chatham}%
\end{thebibliography}

\end{document}